\documentstyle[11pt]{article}
\input FEYNMAN
\title{\vspace{-1.in} \hfill {\small\rm TUM-HEP-310/98} \\
\hfill {\small\rm SFB 375-298} \\~\\
Gap Equation in  Finite-temperature Three-dimensional QED beyond the 
Bare-vertex Approximation}
\author{ George Triantaphyllou\thanks{e-mail:georg$@$physik.tu-muenchen.de}
$\;$ \\{\it Alexander von Humboldt Fellow}\\~ 
\\~  \\{\it Institut f\"ur Theoretische Physik, Technische 
Universit\"at M\"unchen}\\
{\it James-Franck-Strasse, D-85748 Garching, GERMANY }}
\begin{document}
\setlength{\baselineskip}{24pt}
\maketitle
\begin{abstract}
Dynamical mass generation in a three-dimensional version of 
finite-temperature $QED$
is studied with the help of 
Schwinger-Dyson equations in the real-time formalism. 
We go beyond the bare-vertex approximation and include
wave-function renormalization effects. This introduces 
a system of two integral equations which are solved
numerically.   
In order to increase the reliability of the results,    
fermion and photon self-energies varying independently with
energy and momentum are used. 
The method applied enables a detailed study of the behaviour of
the theory with increasing temperature and number of
fermion flavours.  

~
\vspace{2.in}
\end{abstract}
\setcounter{page}{0}
\pagebreak

\section{INTRODUCTION}
Three-dimensional $QED$ can exhibit interesting  non-perturbative 
phenomena like dynamical mass generation \cite{bunch}-\cite{volo}.    
A finite-temperature version of the theory, referred to sometimes as
$\tau_{3}-QED$, has been suggested
to be relevant to high-temperature 
superconductivity \cite{Mavro}. The critical behaviour of the theory
with varying temperature $T$ and number of fermion flavours $N_{f}$ was
recently studied numerically 
in the Schwinger-Dyson formalism \cite{georg}. In that paper, a bare-vertex
approximation was used in the gap equation for the fermion self-energy
$\Sigma$, and wave-function renormalization was neglected. These
approximations correspond to a particular truncation of the 
Schwinger-Dyson hierarchy and to the simplest way of making 
this scheme useful for computations.  
Since this could influence the final results, we return to 
the problem in order to study the impact of these approximations. 
Their potential importance has been stressed in \cite{Penni}, 
especially in connection with the behaviour of the theory when  
the number of fermion flavours is varied. 
This has led to several interesting studies at
zero-temperature trying to include these effects \cite{Nash}-\cite{Maris}, in
order to solve the controversy on whether there is a critical 
number of fermion flavours beyond which there is no mass generation. At 
finite temperature, similar studies have so far used several approximations
that we will relax in this study. 

The first goal of this paper is to provide an estimate  
  for the ratio $r=2M(0,0)/k_{B}T_{c}$ \cite{Mavro} that is free
from severe approximations used frequently in  the past.
Here
$M(0,0)$ is the fermion mass function at zero momentum and energy
and $T_{c}$ is the critical temperature beyond which there is no
mass generation.  Apart from being theoretically very interesting, 
this quantity is of direct physical relevance since it can be
compared to corresponding values measured in certain
high-temperature superconductors. The second  goal   
is to provide 
a reliable phase diagram of the theory with respect to $T$ and $N_{f}$. 
In order to achieve these goals, the bare-vertex approximation is here
relaxed and
wave-function renormalization is taken into account.  

This complicates the study considerably and constitutes a non-trivial
step forward, since instead of having only
one gap equation, a system of two  integral equations has 
to be solved. In particular, the numerical algorithm becomes  
more complicated than before and each iteration requires more computing
time.   
As in \cite{georg} however, we continue  to work 
with fermion and photon self-energies varying independently with  
energy and momentum, something which by itself is a very 
important improvement of similar  investigations in the past
\cite{instant}. We continue
to neglect the imaginary parts of the photon polarization
functions and of the fermion self-energy for simplicity, 
as other studies have done so far \cite{instant}-\cite{ian2}. A more
detailed  treatment would of course involve the full  propagators
\cite{Smilga}, but this is left to future studies due to the limited
numerical accuracy of the present method.

\section{TAKING WAVE-FUNCTION RENORMALIZATION INTO ACCOUNT} 
The study of the fermion self-energy and mass function in a non-perturbative
context is central to our study, and the Schwinger-Dyson formalism will
provide the basis of this investigation. 
The Schwinger-Dyson equation for the fermion self-energy is given by
\begin{equation}
S^{-1}(p) = S_{0}^{-1}(p) 
- e^{2}\int \frac{d^{3}k}{(2\pi)^{3}}
\gamma^{\mu}S(q)\Delta_{\mu\nu}(k)\Gamma^{\nu}(p,q) 
\end{equation}

\noindent 
where $q = p - k$, $e$ is the dimensionful gauge coupling of the theory 
which we will take to be constant throughout this paper, 
$\Delta_{\mu\nu}$ is the photon propagator with $\mu, \nu = 0, 1, 2$, 
$\Gamma^{\nu}$ is the
full photon-fermion vertex, $\gamma^{\mu}$ is a four-dimensional
representation of the $\gamma$-matrices,
$S_{0}$ is the bare fermion propagator, and the
 finite-temperature
fermion propagator in the real-time formalism 
is given by 
\begin{equation} 
S(p) = \left(\left(1+A(p)\right)p\!\!\!/ + \Sigma(p)\right)
\times \left( 
 \frac{1}{\left(1+A(p)\right)^{2}p^{2} + \Sigma^{2}(p)} -
 \frac{2\pi\delta((1+A(p))^{2}p^{2}+\Sigma^{2}(p))}
 {e^{\beta|p_{0}|}+1}\right),  
 \end{equation}  

\noindent where $\beta = 1/k_{B}T$,  
$A(p)$ is the wave-function renormalization function,  
$\delta$ is the usual Dirac function and we have made a rotation to
Euclidean space.
Note that we avoid the matrix form that the propagator has in 
this formalism, 
since the Schwinger-Dyson equation that we have written down  
involves only a one-loop diagram directly,   
so complications due to the field-doubling problem
do not arise \cite{Land}. Moreover, due to the 
broken Lorenz invariance at finite temperature, 
the wave function renormalization
could in principle affect differently the $p_{0}$ and $|\vec{p}|$
propagator components, i.e.
we should replace $(1+A(p))p\!\!\!/$ by 
$((1+A(p))\gamma^{0}+a)p_{0} + (1+B(p))\gamma^{i}p_{i}$ 
with $i = 1, 2$. For simplicity we will restrict ourselves
to situations where $a=0$, which correspond to a zero
chemical potential, and we will work in the 
approximation where $A(p) = B(p)$ also for non-zero 
temperatures as done in similar studies \cite{ian2} for simplicity, 
even though there is
no {\it a priori} justification for this. We will return to this matter
later.

For the vertex  $\Gamma^{\nu}(p,q) = \Gamma_{A}(p,q)\gamma^{\nu}$
we use two different ans\"atze, where
we take $\Gamma_{A}(p,q)$ to be  equal to $1 + A(q)$ or to the symmetrized one
$1+\left(A(p)+A(q)\right)/2$, and 
$A(q)$ is the same wave-function renormalization function
appearing in the fermion propagator. Even though
these vertices do not satisfy {\it a priori}
the Ward-Takahashi identities, they are expected to incorporate the basic
qualitative features of a non-perturbative vertex at zero temperature 
when used in a Schwinger-Dyson context \cite{Miranski}.  
The first choice has been used in a  
finite-temperature case \cite{ian2}, supported by the qualitative 
agreement of the results of this ansatz with the ones obtained by more
elaborate treatments \cite{Penni}. The second vertex is discussed in 
Ref.\cite{Ball} and we will refer to it as the symmetric vertex.
Even though  these 
choices correspond to  specific truncations of the Schwinger-Dyson hierarchy,
they expected to give qualitatively reliable results. 

Moreover, the photon propagator in the Landau gauge
is given by \cite{Mavro} 
\begin{equation}
\Delta_{\mu\nu}(k) = \frac{Q_{\mu\nu}}{k^{2}+\Pi_{L}(k)}
+\frac{P_{\mu\nu}}{k^{2}+\Pi_{T}(k)}
\end{equation} 
\noindent where 
\begin{eqnarray}
Q_{\mu\nu}&=&(\delta_{\mu 0}-k_{\mu}k_{0}/k^{2})\frac{k^{2}}{\vec{k}^2}
(\delta_{\nu 0} - k_{\nu}k_{0}/k^{2}) \nonumber \\ 
P_{\mu\nu}&=&\delta_{\mu i}(\delta_{ij} - 
k_{i}k_{j}/\vec{k}^2)\delta_{\nu j}
\end{eqnarray}

\noindent  with $i, j = 1, 2$, 
and where we neglect its temperature-dependent delta-function
part since it is expected to give a vanishingly small contribution
\cite{georg},  \cite{ian1}, \cite{delref}.
The longitudinal and 
transverse photon polarization functions $\Pi_{L}$ and $\Pi_{T}$
are given in \cite{ian1}, where they
are calculated in a massless-fermion approximation. 
Note that in principle one should also
have an integral equation for the polarization functions coupled to 
$M$ and $A$, but this would complicate this study even numerically
too much, and we therefore continue to work with this approximation.

Identifying the parts of this equation with the same spinor structure, 
we reduce the problem to that of a system of two three-dimensional  
integral equations involving two functions varying independently with 
$p_{0}$ and $|\vec{p}|$.
The equations take the following form:
\begin{eqnarray}
 M(p_{0},|\vec{p}|)&=&\frac{\alpha}{N_{f}(1+
 A(p_{0},|\vec{p}|))}\int 
\frac{dk_{0}|\vec{k}|d|\vec{k}|d\theta}{(2\pi)^{3}}  
\frac{ M(q_{0},|\vec{q}|) R(p_{0},|\vec{p}|,q_{0},|\vec{q}|)}{q^{2}+ 
M^{2}(q_{0},|\vec{q}|)} 
\times \nonumber \\&&\nonumber \\ 
&& \times \sum_{P=L,T} 
\frac{1}{k^{2}+\Pi_{P}(k_{0},|\vec{k}|)}  
\nonumber \\ && \nonumber \\ 
&&-\frac{\alpha}{N_{f}(1+ A(p_{0},|\vec{p}|))} 
\int\frac{|\vec{k}|d|\vec{k}|d\theta}{(2\pi)^{2}}  
\frac{M(E,|\vec{q}|)R(p_{0},|\vec{p}|,E,|\vec{q}|)}
{2E(e^{\beta E} + 1)} 
 \times  \nonumber \\&&\nonumber \\&&\times
 \sum_{\epsilon=1,-1}\sum_{P=L,T} 
\frac{1}{(p_{0}-\epsilon E)^{2}+\vec{k}^{2}+
\Pi_{P}(p_{0}-\epsilon E,|\vec{k}|)}  \nonumber \\&&\nonumber \\ 
A(p_{0},|\vec{p}|)&=&\frac{\alpha}{N_{f}p^{2}}\int 
\frac{dk_{0}|\vec{k}|d|\vec{k}|d\theta}{(2\pi)^{3}}  
\frac{R(p_{0},|\vec{p}|,q_{0},|\vec{q}|)}{q^{2}+ 
 M^{2}(q_{0},|\vec{q}|)} 
\times \nonumber \\&&\nonumber \\  
&&\times \left(\frac{Q(p_{0},\vec{p},k_{0},\vec{k})}{k^{2}+
\Pi_{L}(k_{0},|\vec{k}|)} 
+\frac{P(p_{0},\vec{p},k_{0},\vec{k})}{k^{2}
+\Pi_{T}(k_{0},|\vec{k}|)} \right) 
 \nonumber \\&&\nonumber \\ 
&&-  \frac{\alpha}{N_{f}p^{2}} 
\int\frac{|\vec{k}|d|\vec{k}|d\theta}{(2\pi)^{2}}  
 \frac{R(p_{0},|\vec{p}|,E,|\vec{q}|)}
{2E(e^{\beta E} + 1)} 
 \times  \nonumber \\&&\nonumber \\&&\times \sum_{\epsilon=1,-1}\left( 
\frac{Q(p_{0},\vec{p},p_{0}-\epsilon E,\vec{k})}{(p_{0}-\epsilon E)^{2}
+\vec{k}^{2}+
\Pi_{L}(p_{0}-\epsilon E,\vec{k})} + \right. \nonumber \\&&\nonumber\\ 
&&+\left.
\frac{P(p_{0},\vec{p},p_{0}-\epsilon E,\vec{k})}{(p_{0}-\epsilon E)^{2}
+\vec{k}^{2}+
\Pi_{T}(p_{0}-\epsilon E,\vec{k})}\right), 
\label{eq:fingap}
\end{eqnarray}

\noindent 
where $\alpha = e^{2}N_{f}$, $R(p_{0},|\vec{p}|,q_{0},|\vec{q}|)=
\frac{\Gamma_{A}(p_{0},|\vec{p}|,q_{0},|\vec{q}|)}{1+A(q_{0},|\vec{q}|)}$, 
and  it is more convenient to work with the mass function 
$M(p_{0},|\vec{p}|)=
\Sigma(p_{0},|\vec{p}|)/(1+ A(p_{0},|\vec{p}|))$. The function $R$ is 
obviously equal to the identity for the first vertex choice. 
We also sum over the photon polarizations $P=L, T$ and
over the two roots of the delta function by introducing $\epsilon=1,-1$. 
The quantity $E$ is  approximated by the relation 
$E^{2} \approx |\vec{q}|^{2} + M^{2}(0,0)$ \cite{georg}, where
use of the delta-function property $\delta(ax)=\delta(x)/|a|$ has been
made.

\noindent  Furthermore, the functions $Q$ and $P$ are given by
\begin{eqnarray}
Q(p_{0},\vec{p},k_{0},\vec{k})&=& 2\left(p_{0}-\frac{(pk)k_{0}}{k^{2}}\right)
\frac{k^{2}}{\vec{k}^2}\left(q_{0}-\frac{(qk)k_{0}}{k^{2}}\right)
- pq \nonumber\\
P(p_{0},\vec{p},k_{0},\vec{k})&=& 2\left(\vec{p}\;\vec{q}
-\frac{(\vec{p}\vec{k})(\vec{k}\vec{q})}{\vec{k}^2} \right) -pq. 
\end{eqnarray}
\noindent  One can easily check that 
$Q + P = - 2(pk)(kq)/k^{2}$, which would reproduce the result of 
\cite{ian2} if one takes $\Pi_{L}(k)=\Pi_{T}(k)=\Pi(k)$ and 
switches to imaginary-time formalism (note that our $q$ has the opposite
sign than the one in \cite{ian2}).

The equations for
$M(p_{0},|\vec{p}|)$ and $A(p_{0},|\vec{p}|)$ 
have to be solved self-consistently, and the equation for the mass function 
actually always accepts, apart from the solutions we will seek,
the trivial solution as well. An analytical study of such a system
would not be possible without severe approximations, so we proceed to
the numerical solution of the equations given above.

\section{SOLVING THE SYSTEM OF EQUATIONS}
\subsection{The algorithm}
In order to attack the problem numerically, we have to discretize
our momentum space. The technique will in principle be similar to
the one in \cite{georg}, but the algorithm will be substantially
complicated due to the fact that we now have to solve a system of 
two four-dimensional integral equations involving functions of
two variables.   
The gap equations are both ultraviolet (UV) and infrared (IR) finite, 
with $\alpha$ providing an effective UV cut-off \cite{appel} and the 
fermion mass function providing a physical IR cut-off. 
Therefore, we use UV and IR cut-offs $\Lambda_{IR}$ and
$\Lambda_{UV}$ only for numerical reasons.  
However, we are always careful to keep the temperature $T$ and the 
mass function $M$ within the range of $\Lambda_{IR}$ and 
$\Lambda_{UV}$.  
It is a well-known problem however that the existence of this
IR cut-off does not allow us to draw firm conclusions on the 
criticality of the theory, since a mass function dropping 
below this cut-off could be small but still not exactly zero. We will
return to this issue later.
The complexity of the algorithm does not allow us to consider
$\Lambda_{IR}/\Lambda_{UV}$ ratios smaller than $10^{-4}$.
Since the 
momentum space spans several orders of magnitude, we use logarithmic
variables and discretize the squares of external and loop momenta 
according to $\log_{10}(\Lambda_{IR}^{2})+
\frac{i}{n}\log_{10}{(\Lambda_{UV}^{2}/\Lambda_{IR}^{2})}$,  and the 
integration angle $\theta$ according to $2\pi i/n$, where $i = 1, ..., n$.  
We have therefore a five-dimensional
lattice, with three dimensions coming from the integrals and two
from the external momenta. 

The results presented in this paper 
are for a lattice having 16 points in each  
of the five
dimensions. We checked the stability of our results for lattices of 
other sizes as well, although computing-time limitations did not 
allow us to consider much larger lattices. 
We then use a relaxation method 
described in Ref.\cite{georg}
to solve our system of equations self-consistently by iterations. 
We  do not  exceed a $10\%$ numerical accuracy 
of the results in the worst cases, which we judge as satisfactory
due to the complexity of the algorithm. We trace back this numerical
uncertainty to the minimum achievable
root-mean-square deviation of the input and output
configurations of the algorithm. This deviation is much smaller if one
neglects the high-momentum regions where the solutions are vanishingly
small and where numerical errors appear easier. 
Note however that the mean relative
difference of the input and output configurations 
is monotonically decreasing  with each iteration, like the deviation,
but reaches values as low as $10^{-4}$, which
we consider as an indication of an overall convergence. On the contrary, 
in non-converging cases this
difference is increasing and the configurations tend rapidly to zero, 
apart from the fact that the deviation is oscillating 
randomly taking sometimes unusually large values. 
We therefore have a clear-cut convergence criterium to enable us to 
study the critical behaviour of the theory. 

\subsection{Behaviour of the theory with temperature} 
The solution at $T=0$ and $N_{f}=2$ for the functions
$\Sigma(p_{0},|\vec{p}|)$ and 
$ - A(p_{0},|\vec{p}|)$  for the vertex with $\Gamma_{A}(p,q) = 1+A(q)$ 
is given in 
Figs. 1 and 2 respectively. The general form of 
these functions does not change with increasing temperature, 
varying of $N_{f}$ or by using the symmetric vertex instead. 
The function $\Sigma(p_{0},|\vec{p}|)$ falls 
as expected with increasing momentum, and is of the same form
as the mass function $M(p_{0},|\vec{p}|)$. 
The function $A(p_{0},|\vec{p}|)$ is always in the range
between -1 and 0 as required \cite{Claude}, it is tending to zero
for increasing momenta, and it is of the same
form and magnitude as the approximate form used in \cite{ian2}. 
In view of the approximation
$A(p)=B(p)$ for the wave-function renormalization stated at the beginning, 
the roughly symmetric behaviour of the two functions $A$ and $\Sigma$ with 
respect to $p_{0}$ and $|\vec{p}|$
argues in support for the self-consistency of the calculation. Moreover,
the low-momentum values of the wave-function renormalization,
which are on the order of $A(0,0) \approx -0.3$, indicate  the 
importance of including its effects in a complete calculation in this
gauge.

\begin{figure}[p]
\vspace{4.5in}
\includegraphics{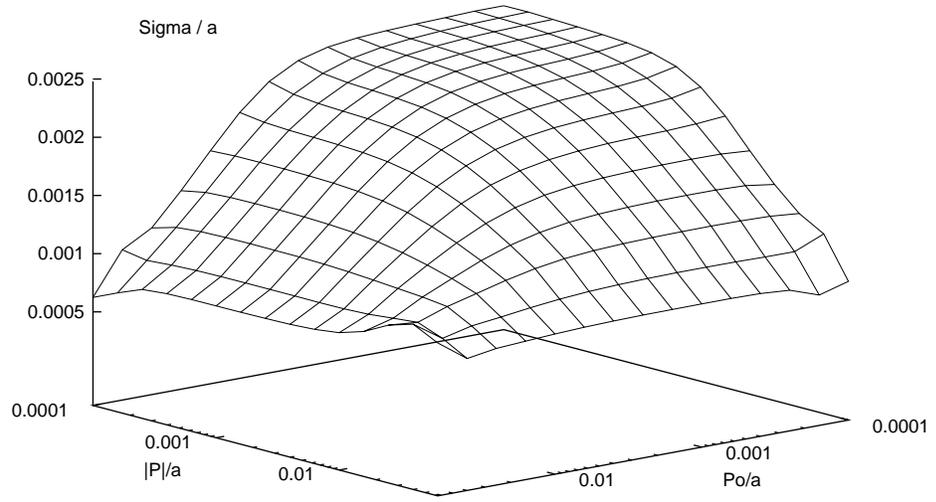}
\caption{The fermion self-energy $\Sigma(p_{0},|\vec{p}|)$ 
 at zero temperature and for $N_{f} = 2$,
$\Lambda_{UV}/\alpha = 0.1$, as a function of energy and momentum in
logarithmic scale. All quantities are scaled by $\alpha$.}  
~\\
\label{fig:fig1}
\end{figure}

\begin{figure}[p]
\vspace{4.5in}
\includegraphics{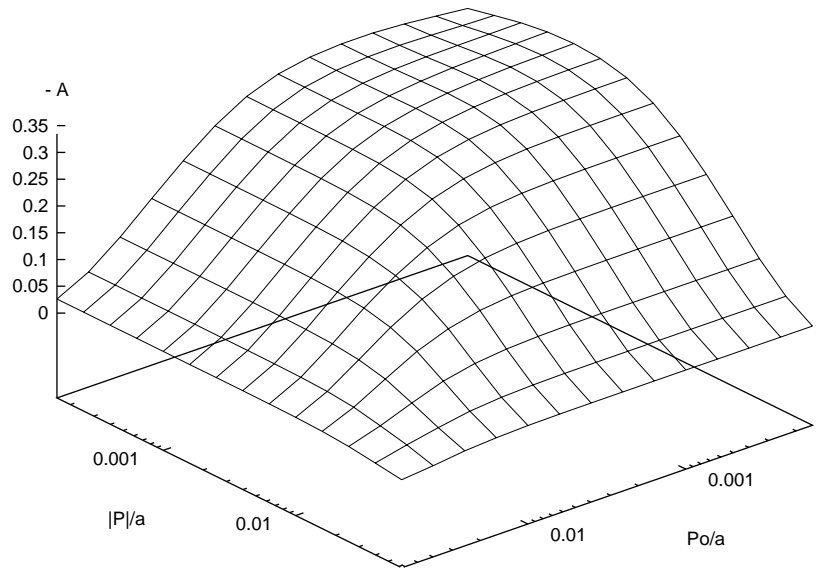}
\caption{The opposite of the  
wave-function renormalization $-A(p_{0},|\vec{p}|)$ at zero temperature
and for $N_{f} = 2$, $\Lambda_{UV}/\alpha = 0.1$.}  
~\\
\label{fig:fig2}
\end{figure}

For a given number of fermion flavours $N_{f}$, when the temperature
exceeds some critical value there is no solution for the fermion mass function 
but the trivial one. We infer  that not only by the fact that the 
relaxation algorithm does not converge, but also because the output 
configurations tend rapidly to zero and after a number of iterations 
they fall below the IR cut-off. 
One should note that the temperature at which this happens could be
somewhat smaller than the critical one, since the zero-momentum
mass function at slightly
smaller temperatures  is not very much smaller than the one at zero
temperature.
The abrupt fall of the
solution with temperature 
thus could possibly be an indication that the initial smooth input
configuration is
unable to converge to the correct solution, which is rapidly 
varying but still not zero.
Since the dimensionless
ratio $r$ of (twice)
the zero-momentum and zero-temperature fermion mass function 
divided by
this critical temperature is important for the study of solid-state 
systems, we list in Tables 1 and 2 its values for various 
choices of $N_{f}$ and 
$\Lambda_{UV}/\alpha$, for our two different choice of vertices. 
Note that the case $N_{f} = 2$ is the one 
relevant to high-temperature superconductivity.

   \begin{table}[p]
    \begin{tabular}{||c
 ||@{\hspace{2mm}}c@{\hspace{2mm}}|@{\hspace{2mm}}c@{\hspace{2mm}}
 |@{\hspace{2mm}}c@{\hspace{2mm}}|@{\hspace{2mm}}c@{\hspace{2mm}}
 |@{\hspace{2mm}}c@{\hspace{2mm}}|@{\hspace{2mm}}c@{\hspace{2mm}}
 |@{\hspace{2mm}}c@{\hspace{2mm}}|@{\hspace{2mm}}c@{\hspace{2mm}}
 |@{\hspace{2mm}}c@{\hspace{2mm}}||}  \hline
\rule[-3mm]{0cm}{8mm}
Fermion flavours $\Rightarrow$
&\multicolumn{3}{c||@{\hspace{2mm}}}{$N_{f}=1$} &  
 \multicolumn{3}{c||@{\hspace{2mm}}}{$N_{f}=2$} & 
 \multicolumn{3}{c||}{$N_{f}=3$}  
  \\  \cline{1-10} 
\rule[-3mm]{0cm}{8mm}
 $ \Lambda_{UV}/\alpha \; \Downarrow$ 
&$m_{o}$&$t_{c}$&$r$&$m_{0}$&$t_{c}$&$r$&$m_{0}$&$t_{c}$&$r$ \\ \cline{1-10}
\rule[-3mm]{0cm}{8mm}0.1 &21&4.5&9.4&3.3&0.6&11.0&0.65&0.12&10.6  
\\ \cline{1-10}
\rule[-3mm]{0cm}{8mm}0.5 &32&6.2&10.4&4.0&0.72&11.1&1.3&0.26&10.0  
\\ \cline{1-10}
\rule[-3mm]{0cm}{8mm}1   &32&6.1&10.5&4.4&0.75&11.7&1.4&0.27&10.4 
\\ \hline \hline 
      \end{tabular}  
 \caption{ The quantities $m_{0}=10^{3} \times M(0,0)/\alpha$
 at $T=0$, $t_{c}= 10^{3} \times k_{B}T_{c}/\alpha$ and 
 $r=2m_{0}/t_{c}$ for different values of  
flavours and ultra-violet cut-offs $\Lambda_{UV}$. The vertex
with $\Gamma_{A}(p,q)=1+A(q)$ is used.}  
~\\
\label{table:rhota2}
     \end{table}

   \begin{table}[p]
    \begin{tabular}{||c
 ||@{\hspace{2mm}}c@{\hspace{2mm}}|@{\hspace{2mm}}c@{\hspace{2mm}}
 |@{\hspace{2mm}}c@{\hspace{2mm}}|@{\hspace{2mm}}c@{\hspace{2mm}}
 |@{\hspace{2mm}}c@{\hspace{2mm}}|@{\hspace{2mm}}c@{\hspace{2mm}}
 |@{\hspace{2mm}}c@{\hspace{2mm}}|@{\hspace{2mm}}c@{\hspace{2mm}}
 |@{\hspace{2mm}}c@{\hspace{2mm}}||}  \hline
\rule[-3mm]{0cm}{8mm}
Fermion flavours $\Rightarrow$
&\multicolumn{3}{c||@{\hspace{2mm}}}{$N_{f}=1$} &  
 \multicolumn{3}{c||@{\hspace{2mm}}}{$N_{f}=2$} & 
 \multicolumn{3}{c||}{$N_{f}=3$}  
  \\  \cline{1-10} 
\rule[-3mm]{0cm}{8mm}
 $ \Lambda_{UV}/\alpha \; \Downarrow$ 
&$m_{o}$&$t_{c}$&$r$&$m_{0}$&$t_{c}$&$r$&$m_{0}$&$t_{c}$&$r$ \\ \cline{1-10}
\rule[-3mm]{0cm}{8mm}0.1 &22&4.6&9.6&3.4&0.67&10.1&1.1&0.21&10.5     
\\ \cline{1-10}
\rule[-3mm]{0cm}{8mm}0.5 &29&5.7&10.2&4.7&0.85&11.1&1.4&0.27&10.4   
\\ \cline{1-10}
\rule[-3mm]{0cm}{8mm}1   &31&5.8&10.7&5.0&0.9&11.1&1.5&0.28&10.7 
\\ \hline \hline 
      \end{tabular}  
 \caption{ The same quantities as Table 1 
using the symmetric vertex.}  
~\\
\label{table:rhota3}
     \end{table}

\begin{figure}[p]
\vspace{4.5in}
\includegraphics{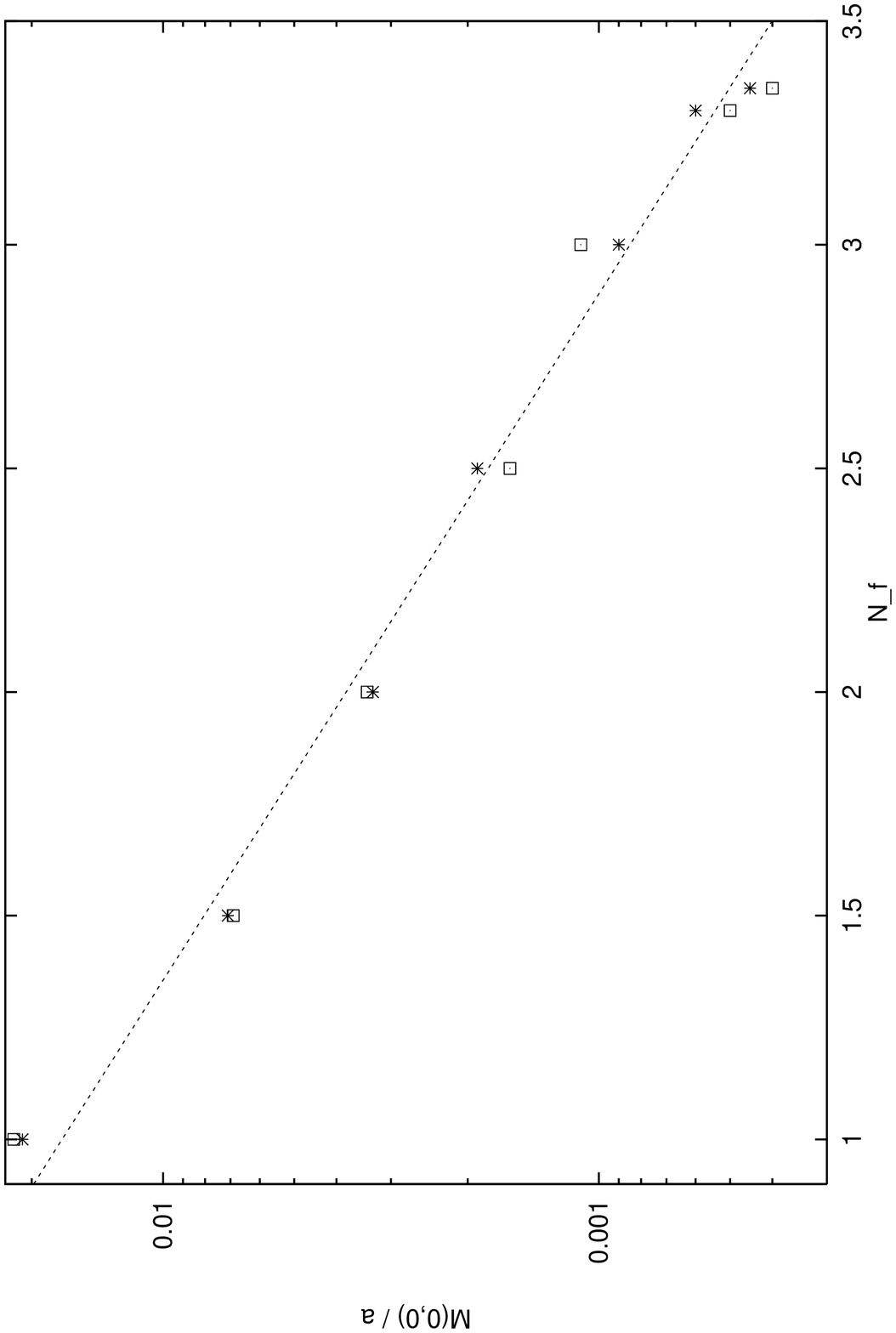}
\caption{The fermion mass function at zero momentum and zero temperature, 
scaled by $\alpha$ and on a logarithmic scale,  
with respect to
$N_{f}$ for a ratio $\Lambda_{UV}/\alpha = 0.1$. 
Stars and squares correspond to the vertex  with  $\Gamma_{A}=1+A(p)$ and the 
symmetric vertex respectively.
We fit our results for both vertices with the curve 
$M(0,0)/\alpha = e^{-1.5N_{f}}/13.1$.   
 Values of $N_{f}$ larger than 3.35 are not
considered, because then the self-energy falls below the IR-cut-off.}  
~\\
\label{fig:fig3}
\end{figure}

The fermion mass function falls with increasing momenta  and the results 
obtained should not depend on the choice of the UV cut-off. Even though
the limited
number of lattice points  did not allow us to explore the
region $\Lambda_{UV} > \alpha$,  this is not expected to
be physically interesting
since the critical temperature and the fermion mass function is always
 orders of magnitude smaller than $\alpha$. For the same reason, 
 we tend to trust more our results with lower $\Lambda_{UV}/\alpha$
ratios, since then the lattice spacing becomes smaller and the integrations
more accurate. At momenta of order $\Lambda_{UV} = 0.1 \alpha$ the
mass function is already negligibly small, and thus our results
are not influenced by such a UV cut-off. We were not able to consider
even smaller $\Lambda_{UV}/\alpha$ as in \cite{georg} due to numerical
problems caused by the complexity of the algorithm.

We note that in general the value of $r$ remains roughly stable for 
various choices for
$N_{f}$ and $\Lambda_{UV}$. It is in fact  more stable than
the $r$ ratio computed in \cite{georg}, which could be an indication
that the wave-function renormalization $A(p_{0},|\vec{p}|)$ 
plays a -at least numerically- stabilizing role 
in the system under study. 
In this respect, the results presented here could be more trustworthy. 
Furthermore, comparing the Tables 1 and 2 one can see that the $r$-ratio
is not considerably affected by the particular choice of our vertices. 
Moreover, we note that its value is
concentrated approximately around $r \approx 10.7$, which is 
comparable to the values obtained in \cite{georg}  
which neglected $A(p_{0},|\vec{p}|)$.
The ratio $r$ found is also comparable to  
typical values obtained in Ref. \cite{ian2}, which includes wave-function
renormalization effects but uses several approximations which we 
were able to by-pass in this study. We confirm therefore that adding 
these effects does not influence the behaviour of the theory in a
significant way.
We have  to note moreover
that our $r$ values are somewhat larger than
the value $r \approx 8$ measured for
some high-temperature superconductors \cite{Schles}.
However, as already discussed, we could be overestimating this ratio 
because of a possibly poor convergence of the algorithm for
temperatures close to the critical one.  

\subsection{Behaviour of the theory with the number of fermion flavours}
At zero-temperature, this theory is known to exhibit also an interesting
behaviour with the number of fermions $N_{f}$. 
In Fig. 3 we plot the zero-momentum and zero-temperature fermion
mass function with respect to  $N_{f}$, fitted with the same  
exponential curve for both vertices, since their behaviour is 
similar. There are two reasons why we do
not try to fit them with a function non-analytic in $N_{f}$, as was done
in \cite{georg}. One reason is that
the non-analytic form used before and predicted by some studies is expected 
to be valid only for a 
number of fermion flavours close to a possibly
critical value $N_{c}$, so efforts to use
the same functional
form for $N_{f}$ away from this value could give a misleading
value for $N_{c}$. 
The other is that our present results for the ratio
$r$ seem to be quite stable with $N_{f}$, and we would like to be able to
estimate in a simple way
some sort of ``mean" $r$-ratio by combining these data with the
ones showing the $T_{c}$ dependence on $N_{f}$, as will be described below.

At $N_{f} \approx 3.35$, the mass function is still roughly four 
times larger than the cut-off. 
When $N_{f} ~^{>}_{\sim}\; 3.35$,  
our algorithm  does not converge and the mass function 
tends fast below  the IR cut-off. Note  that this value did not
vary substantially for different choices of $\Lambda_{IR}$ or the
vertex. 
This behaviour
could indicate that $N_{f} \approx 3.35$  is some critical point beyond
which dynamical mass generation is impossible. 
However, numerical limitations did not allow us to explore the region 
$\Lambda_{IR} < 10^{-4} \Lambda_{UV}$. 
In other words, if 
there were a solution for the mass function smaller than roughly 
$10^{-4}\Lambda_{UV}$, our algorithm would not be able to find it because it
would fall below the IR cut-off. The finite size of our lattice therefore 
does not allow us to draw firm conclusions on the matter, and the
fitting curve should not be extrapolated 
beyond this limiting $N_{f}$ value. 

The value of $N_{f}$ where this behaviour comes in effect
is  remarkably close to the one
quoted in the numerical study of \cite{Maris}, and it is also close to
our previous result \cite{georg}.
These values are
however roughly $25\%$ smaller than the one quoted in the
theoretical studies of Refs. 
\cite{Kondo}, \cite{ian3} which take wave-function 
renormalization effects into account.
A  similar study \cite{Nash}, 
which includes a calculation of the
fermion-field anomalous dimension to 
second-order in $1/N_{f}$, 
predicts a critical  value 
$N_{f}  \approx 3.28$, which 
is only slightly larger than the 
theoretical prediction neglecting wave-function 
renormalization which gives 
 $N_{f} = \frac{32}{\pi^{2}} \approx 3.24$ \cite{appel}, and  
 quite close to the value  where our algorithm looses convergence.

\begin{figure}[p]
\vspace{4.5in}
\includegraphics{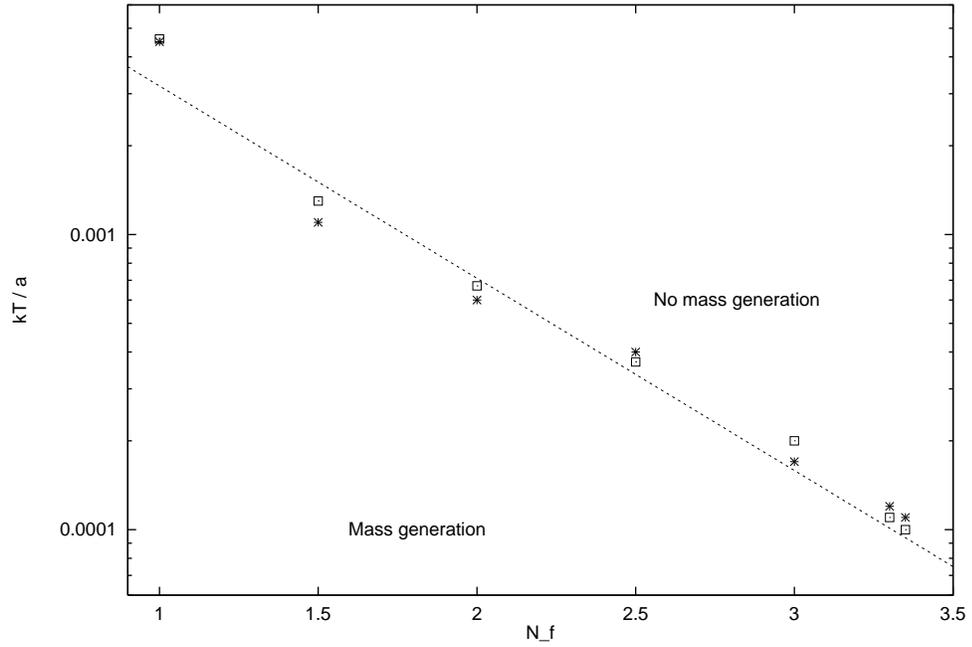}
\caption{The  
phase diagram of the theory with respect to the temperature scaled by 
$\alpha$ and in
logarithmic scale and
the number of fermion flavours, for a ratio
$\Lambda_{UV}/\alpha = 0.1$. 
Stars and squares correspond to the vertex  with  $\Gamma_{A}=1+A(p)$ and the 
symmetric vertex respectively.
We fit the data points with the
curve $k_{B}T/\alpha = e^{-1.5 N_{f}}/70$. This curve should not
be extrapolated for $N_{f} ~^{>}_{\sim}\; 3.35$.}  
~\\
\label{fig:fig4}  
\end{figure}

In Fig. 4 we plot the phase diagram of the theory with respect to 
$N_{f}$ and $k_{B}T$. It separates two regions of the parameter space
which either allow or do not allow dynamical mass generation.  
The approximate 
stability of $r$ with increasing $N_{f}$ lead us to rescale the 
data points of this plot and fit these simultaneously with the
data points of Fig. 3, with the same exponential curve $e^{-1.5N_{f}}/13.1$
for both vertices, since they exhibit a similar behaviour. 
The slope of this exponential fits also relatively well the data  on the two
 plots separately, even though the slope of the exponential of 
 the mass function separately 
 is closer to -2. 
It is however lower in absolute value than the one reported in 
\cite{Miranski} . We note also that the fall of the critical temperature 
 with the number of fermion flavours is
 slower than in \cite{georg}, which was found to be -2.5. 
 The fast fall in that reference was the
 cause of the increase of the $r$-ratio with $N_{f}$, and we interpret it
 as poor convergence of that algorithm with large $N_{f}$ values,
 because then the self-energy becomes smaller and more vulnerable
 to instabilities coming from  numerical errors. 
The (double of the) rescaling factor  was found to be
$\bar{r} =2M(0,0)/k_{B}T \approx 10.7$ and provides some kind of 
``mean" $r$-ratio which is independent of $N_{f}$. 
Its approximate value could actually be guessed by
the data in Tables 1 and 2. 

The choice of an exponential fitting curve was only made to describe
``phenomenologically"
the general tendency of the data and to provide a measure  for a
$r$-ratio independently of $N_{f}$, and is reminiscent of the results
in Ref. \cite{Penni} but with a steeper slope. 
However, there are also studies that predict a
non-analytic behaviour of $\Sigma$ for $N_{f}$ near its critical
value \cite{appel}. One could speculate that
such a behaviour comes into effect also for our larger $N_{f}$ data, since 
the slope of a curve connecting the large-$N_{f}$
points of both Fig. 3 and 4 is larger than the one of the
general fit. Moreover, as already noted, near these $N_{f}$ values
the mass function is still
roughly a factor of four 
larger than the IR cut-off, so the value of $N_{f}$ quoted cannot be 
easily argued to be just an artifact of the finite lattice size. 
Lack of convergence of the algorithm and
fall of the mass function below the IR cut-off  however do not allow us to
test the precise behaviour of the theory near $N_{f} \approx 3.35$, and
it is also clear that the fitting curves should not be extrapolated for 
$N_{f}$ larger than this value. 

We remind the reader that the 
results  quoted above
were  obtained with a photon propagator in the Landau gauge,
which is convenient and widely used for this type of calculation. 
In addition, it possibly enables the wave-function renormalization to play a
stabilizing role in the numerical algorithm and provide more
accurate results.
Since the final answers should be gauge-invariant however, 
one should in principle check
whether other gauges, with corresponding
vertices consistent with the Ward-Takahashi identities,
give the same answer.  In particular,
 calculations in the non-local gauge \cite{nonlo} eliminate the wave-function
renormalization dependence of the results, since $A(p) = 0$,
which means that the ratio $r$ or the possibly critical number of 
fermion flavours should not 
be influenced by $A(p)$. In such a case, however, a different vertex
choice than ours should be made, so that it does not reduce to the bare one
when $A(p) = 0$. In our study, therefore, it is the taking into account of the 
wave-function renormalization in conjunction with a non-bare vertex  
that could potentially lead to results different from \cite{georg}. 

\section{CONCLUSIONS}
To conclude, we were able to 
solve a system of two integral equations for the
fermion mass function and wave-function renormalization for
 a finite-temperature version of three-dimensional $QED$,   
by applying a numerical relaxation technique.  
The results presented are of course specific to the dimensionality of 
the system under study and to the abelian character of the theory. 
Both functions as well as the photon self-energies are taken
to be energy- and momentum-dependent, use of two different
non-bare vertices is 
made, and this leads to a more
accurate and detailed study of the criticality of the 
theory.  
One main result is a $r$-ratio of about 10.7, which 
is close to previous numerical studies, confirming that including 
wave-function renormalization, in conjunction with two different 
non-bare fermion-photon vertices,  does not affect the theory in a
significant way.
The other important result is the
existence of a possibly critical fermion 
flavour number of roughly 3.35, which
is  also consistent with some theoretical expectations and 
other numerical calculations. 
But the use of an IR cut-off does not
allow the precise exploration of the possible criticality at
this stage. 

It should be stressed here that such an agreement could not
be taken for granted {\it a priori}, in view of the seriousness of 
truncations and approximations used in the past. It is the first time
that such a study drops most of the approximations applied so far
and allows a reliable description of the behaviour of the theory. 
We estimate the numerical uncertainty for 
the values quoted, which comes mainly from the
convergence criteria imposed, at about $\pm 10\%$. 
The specific truncations of the Schwinger-Dyson hierarchy chosen in 
this work could of course easily introduce an additional 
theoretical error of comparable magnitude. 
We also note that a  larger 
lattice than the one we used could be expected to improve the
convergence of the algorithm and help us understand 
the behaviour of the fermion mass function at regions  
even closer to the critical temperature and flavour number.
Furthermore, it remains to
be seen if  the inclusion of the imaginary parts of the self-energies 
could influence these results considerably.

\noindent {\bf Acknowledgements} \\
I thank B. Bergerhoff for useful discussions and
N. Mavromatos for bringing relevant calculations  in the
non-local gauge to my attention. 
This research is supported by an {\it Alexander von Humboldt Fellowship}.

\end{document}